\documentclass[]{article}
\usepackage{authblk}
\usepackage[title]{appendix}
\usepackage{color}
\usepackage[dvipsnames]{xcolor}
\usepackage{amssymb}
\usepackage{amsmath,mathrsfs}
\usepackage{braket}
\usepackage{bm} 

\usepackage[ruled]{algorithm2e}
\usepackage{algorithmic}
\usepackage{setspace,soul,comment}
\usepackage{graphicx}
\usepackage{grffile}
\usepackage{subfigure}
\usepackage{hyperref}

\usepackage{url}
\usepackage{cleveref}
\usepackage[utf8]{inputenc}
\usepackage{geometry}
\usepackage{float}
\usepackage[flushleft]{threeparttable}

\usepackage{footnote}

\usepackage{colortbl}

\title{Mutation frequency time series reveal complex mixtures of clones in the world-wide SARS-CoV-2 viral population}

\author[1]{Hong-Li Zeng}
\author[1]{Yue Liu}
\affil[1]{School of Science, Nanjing University of Posts and Telecommunications, Nanjing 210023, China}

\author[2,3]{Vito Dichio}
\affil[2]{Inria Paris, Aramis Project Team, Paris, France}
\affil[3]{Institut du Cerveau, ICM, Inserm U 1127, CNRS UMR 7225, Sorbonne Université, Paris, France}

\author[4]{Kaisa Thorell}
\author[4 $^\ast$]{Rickard Nord\'en }
\affil[4]{Department of Infectious Diseases, Institute of Biomedicine \authorcr Sahlgrenska Academy of the University of Gothenburg, Gothenburg, Sweden}

\author[5 $^\ast$]{Erik Aurell}
\affil[5]{AlbaNova University Center, SE-106 91 Stockholm, Sweden}

\begin{document}
\maketitle
\begin{abstract}
We compute the allele frequencies of the alpha (B.1.1.7), beta (B.1.351) and delta (B.167.2) variants of SARS-CoV-2 from almost two million genome sequences on the GISAID repository. We find that the frequencies of a majority of the defining mutations in alpha rose towards the end of 2020 but drifted apart during spring 2021,
a similar pattern being followed by delta during summer of 2021. 
For beta we find a more complex scenario with frequencies of some mutations rising and some remaining close to zero. 
Our results point to that what is generally reported as single variants is in fact a collection of variants with different genetic characteristics.
For all three variants we further find some alleles with a 
clearly deviating time series.
\end{abstract}


 $^\ast$ The authors made equal contribution to the work and to whom correspondence should be addressed; E-mail:  rickard.norden@microbio.gu.se, eaurell@kth.se

\section{Introduction}
A typical goal of an evolutionary model would be to predict the fixation or extinction in the population of a mutation in the genotype of the individuals, which entails reaching a sufficient understanding of 1. the map genotype-phenotype and 2. the interactions between the phenotype and the surrounding environment: both of them are formidable tasks and object of cutting-edge research at the intersection between physics and biology \cite{neher2011,neher2018,lassig2017,dichio2021}.\\
Any theoretical and modelling effort in the field of population genetics is notoriously hampered by the simple and inevitable lack of experimental data (with some famous exceptions e.g. \cite{good2017,lenski1991}), the major reason being the difficulty in observing evolutionary time-scales under experimental conditions. \\
Suitable data for these kind of models must necessarily refer to rapidly evolving populations, such as bacteria or viruses, whose analysis can lead to crucial predictions of their evolutionary fate. Nevertheless, even when data are available, e.g. as a database of all the genotypes in a viral population, it may be tough to extract useful information from the deafening noise of sequencing errors, finite-size population effects etc.\\
The importance of similar considerations has become evident well beyond the boundaries of academic research in the last year and a half, when the prediction of spread or extinction of a variant of the SARS-CoV-2 has become a matter primarily concern for world-wide public health.

Indeed, following the outbreak of the pandemic, an unprecedented effort has been put in collecting genomic sequences of the SARS-CoV-2 in a public shared repository, in order to boost the research on the virus. The GISAID repository~\cite{GISAID} contains an increasing collection of SARS-CoV-2 whole genome sequences, and has been used to identify mutational hot spots ~\cite{Pacchetti2020} and to infer epistatic fitness parameters~\cite{Zeng2020,Cresswell2021}.
In the last year, Nature has performed an experiment in the growth of Variants of Concern (VOC),
in Pangolin classification referred to as
B.1.1.7, B.1.351 and B.1.617.2. These variants of the virus are today also referred to as 
``alpha", ``beta" and ``delta": earlier they were called
``UK-variant", ``South Africa-variant" and ``Indian variant", as they were first identified in south-east England~\cite{B.1.1.7}, South Africa~\cite{Tegally2021,VoC-6} and India~\cite{Indian_variant}, respectively.
Although most PCR-assays for these variants are based on variability at a few positions, 
full definitions contain more loci, both within and outside the Spike-protein. 
This holds both for Pangolin definitions and others, as we will show.
The frequencies of the mutations at the different positions hence give information on whether these variants in fact grow as large clones, or if they have mutated or recombined into several clones, or if they were several clones from the beginning. 

In this paper, we find that different scenarios hold for the three variants alpha, beta and delta.
This illustrates that allele frequency analysis is a method that can complement phylogenetic approaches in order to more accurately define VOCs at an earlier stage.
The overall message of this work is that comparatively simple time series
analysis can reveal properties of large and complex data which is often
obscured in other methods. In our case this is for instance so for mutations obviously 
mis-labelled in the original definitions. We will see there are quite a few of these.
Our analysis hence emphasizes again the importance of time variability in
complex systems~\cite{Holme2012,Holme-2019,Masuda-2016}, and of time series analysis as a tool to
elucidate relationships in complex data \cite{Marwan2007,Donges2015,Runge2019}.

The paper is organized as follows. In Section~\ref{sec:data}
we describe how we use the uniquely rich data source of GISAID, with the
hope to encourage also further studies in the complex systems community.
The allele frequency analysis at individual mutated locus as well as the joint ones for three Variant of Concern are described in Section~\ref{sec:allele_fre_analysis_annotation}. The annotations of these mutations are also listed in this section. The main results and discussion are given in Section~\ref{sec:results} and~\ref{sec:discussion} respectively.
A preliminary version of this work covering only variants alpha and beta
and using data on GISAID up to early 2021 was deposited on the bioRxiv
pre-print server in April 2021 \cite{earlier-version}.

\section{Data preparation and transformation}
\label{sec:data}
We analyzed the consensus sequences of SARS-CoV-2 deposited in GISAID database (https://www.gisaid.org) ~\cite{GISAID}. These high quality sequences with full lengths (number of bps $\approx 30,000 $) were obtained through the ``complete" and ``high coverage" options on the GISAID interface. GISAID metadata
includes both ``collection time" and ``submission time" of the genome sequences.
As the first is usually two weeks or more earlier than the second
we here used the ``collection time"  option. 

\subsection{Data Collection} 
The data set prepared here were downloaded from GISAID website installments at different times
until the middle of August 2021, with ``collection time" of the genomic sequences until the end of July 2021. In this manner $1,845,260$ genomic sequences are obtained in total which are all included in the allele frequency analysis. 
For deposition, sequences collected in 2020 are packed in one file while the sequences downloaded in 2021 are packed monthly as number of sequences has grown dramatically. All separate data sets are available on the Github repository~\cite{Zeng-github}.  They are also stored physically in a desktop computer with 64G RAM named ``hlz" at Nanjing University of Posts and Telecommunications (NJUPT).  The allele frequencies and the visualizations were both done using MATLAB software on this machine.

\subsection{Multiple-Sequence Alignment (MSA)} 
Multiple sequence alignments (MSAs) were constructed from the downloaded raw sequences with the online alignment server MAFFT~\cite{Katoh2017,Kuraku2013} with the reference sequence ``Wuhan-Hu-1"  carrying Genbank accession number ``MN908947.3". The length of sequences are kept the same as the reference (29903) during sequence alignment.  

An MSA is a big matrix $\mathbf{S}=\{s_i^n|i=1,...,L, n = 1,...,N\}$, composed of $N$ genomic sequences which are aligned over $L$ positions. 
Each entry $s_i^n$ of matrix $\mathbf{S}$ is either one of the 4 nucleotides (A,C,G,T), or ``not known nucleotide'' (N), 
or some minorities,
or the alignment gap `-', introduced to treat nucleotide deletions and insertions. In the main analysis we transformed the minorities `KFY...' into one overall symbol `N' when constructing the MSA. We hence retain six states, i.e., `-NACGT'. We note that with minorities, `-' and `N', deletions can also be detected from the temporal analysis of the allele frequencies (data not shown).

The whole MSA could be divided into pieces along the horizontal direction further for storage. Then each piece is a sub-structures of the entire gene (NSP1 to NSP16, Spike, ORF3a etc) over all pooled time. This operation could reduce the memory requirement during the computation.

\section{Allele frequency of three Variants of Concern (VOC)}\label{sec:allele_fre_analysis_annotation}
 The work focused on the individual and joint allele frequencies analysis for the mutations or deletions listed for B.1.1.7 \cite{B.1.1.7} also known as alpha or ``UK variant", B.1.351 \cite{VoC-6} also known as beta or ``South Africa variant" and B.1.617.2 \cite{Delta_variant} also known as delta or ``Indian variant". 
 
 \subsection{Allele frequency calculation}
 In the time period $\Delta t$, the individual frequencies of a certain nucleotide $x$ at the $i$-th locus are computed by eq.~\eqref{eq:fre_cal}. 
 \begin{equation}\label{eq:fre_cal}
     f_i(x,\Delta t) = \frac{n_i(s,\Delta t)}{N_i(\Delta t)},
 \end{equation}
 With $s\in\{-,N,A,C,G,T\}$ and $\Delta t$ the time length of the analyzed snapshots $n_i(s,\Delta t)$ denotes the number of allele $s$ at locus $i$ during the period of $\Delta t$ while the denominator is the total number of the nucleotides on this locus during the same period $\Delta t$.
 
 Wild type in biology denotes the original variant or the most common variant. 
 The allele frequency analysis with the ``wild type" denoting ``the most common variant" in all
 the pooled data for B.1.1.7 and B.1.351 variant was performed in~\cite{earlier-version}, which is a  proxy for the original variant of SARS-CoV-2 after it transited into its
 human host. 
 In this work we have taken it to be the original variant, which is thus the same
 as the reference sequence Wuhan-Hu-1. A mutant is a deviation from this wild type.
 To visualize if sequences grow together we therefore plot (a) the joint 
 frequency of wild type alleles as a group of loci and (b) the joint frequency of
 the second most prevalent alleles provided by the definitions of three VOCs as shown in the ``mutation locus" column of Table~\ref{tab:B.1.1.7},~\ref{tab:B.1.351} and ~\ref{tab:B.1.617.2}. The relevant formulae are
 
\begin{equation}\label{eq:joint_fre_cal}
     F(v,\Delta t) = \frac{n(v,\Delta t)}{N(\Delta t)},
 \end{equation}
where $v$ is the number of the sequence containing all the corresponding mutations with the B.1.1.7 / B.1.351 / B.1.617.2
while the denominator is the total number of sequences during the same period $\Delta t$.
 
To take into account the effect of evolution time for SARS-CoV-2 virus, both the individual and joint allele frequencies are computed on the time scale of each month from the initial outbreak of the COVID-19 pandemic. 
 There are two obvious outliers denoted by the black square line (16176) and the black dot line (26801) in Fig.~\ref{fig:frequeces_of_loci_monthly_UK}, which have different temporal behaviors compared with the others. In~\cite{earlier-version} we concluded that the mutation of $16176$ was most likely mislabeled. The $23063$ mutation of B.1.351 in  Fig.~\ref{fig:frequeces_of_loci_monthly_SA}  has a distinct pattern from other alleles.

\subsection{Annotated nucleotide mutations}

Differently from the allele frequencies based on the data snapshots (Months), the annotated nucleotide mutations are obtained from the whole prepared data set. With the sorted allele frequencies computed from the whole data set, the most prevalent nucleotide and the second most one are selected as the first (denoted as ``Allele 1") and second allele (``Allele 2"), see in the forth and fifth column of Table~\ref{tab:B.1.1.7}, \ref{tab:B.1.351} and \ref{tab:B.1.617.2} for B.1.1.7, B.1.351 and B.1.617.2 respectively.  Meanwhile the first column is the corresponding gene sub-structures of the defined mutations (second column) as reported in the literature; finally,  the amino acid mutations are shown in the third column in each table.

\subsection*{Definition of B.1.1.7 (``UK variant")} 
In this work we have used the definition of
SARS-CoV-2 Variant of Concern 202012/01
(B.1.1.7) as originally given in
"Technical briefing 1"~\cite{B.1.1.7}  
Table 1 and text above Table 1
(publication date December 21, 2020).
This information with annotations is
given as Table~\ref{tab:B.1.1.7} below. 
All non-synonymous amino acid mutations are listed on the later Pangolin description~\cite{PANGO_UK}.

In a later report from
the same group (Technical 
briefing 6, publication date February
13, 2021~\cite{VoC-6}) another definition of 
B.1.1.7 was given in Table 2a.
That definition differs from the one
used here in that mutation $C28977T$
in the $N$ gene and the six synonymous
mutations have not been retained.

\begin{table}[!ht]
\centering
\caption{Defining mutations for variant B.1.1.7}
\begin{threeparttable}
\begin{tabular}{lllll}
gene sub-structure  & mutation locus \tnote{a} & amino acid mutation &  Allele 1 \tnote{b}  & Allele 2 \tnote{b} \\
\hline
NSP2& C913T & ---      &  C & T \\
NSP3& C3267T & ORF1ab: T1001I &  C & T \\
NSP3& C5388A & ORF1ab: A1708D &   C & A \\
NSP3& C5986T & ---     &   C & T \\
NSP3& T6954C & ORF1ab: I2230T  &   T & C\\
NSP6& 11288-11296 & SGF 3675-3677 del &  ? & - \quad \tnote{*} \\
NSP12& C14676T & ---     &  C & T \\
NSP12& C15279T & ---     &   C & T\\
NSP12& C16176T  & ---     &  \cellcolor{lightgray}T & \cellcolor{lightgray}C\quad\tnote{c}\\
Spike& 21765-21770 & HV 69-70 del     &  ? & -\\
Spike& 21991-21993 & Y144 del     &  ? & -\\
Spike& A23063T & S: N501Y      &  A & T \\
Spike& C23271A & S: A570D     &  C & A\\
Spike& C23604A & S: P681H     & \cellcolor{lightgray}A & \cellcolor{lightgray}C\quad\tnote{c}\\
Spike& C23709T & S: T716I     &  C & T\\
Spike& T24506G & S: S982A     &  T & G \\
Spike& G24914C & S: D1118H     &  G & C\\
M& T26801C & ---     &  \cellcolor{lightgray}C & \cellcolor{lightgray}G\quad\tnote{d}   \\
ORF8& C27972T & ORF8: Q27stop     &  C & T \\
ORF8& G28048T & ORF8: R52I     &  G & T \\
ORF8& A28111G & ORF8: Y73C     &  A & G\\
N& G28280C & N: D3L     &  G & C\quad\tnote{e}\\
N& A28281T & N: D3L     &  A & T\quad\tnote{e}\\
N& T28282A & N: D3L     &  T & A\quad\tnote{e}\\
N& C28977T & N: S235F     &  C & T\\
\hline
\end{tabular}
\begin{tablenotes}
\item[] \footnotesize{The `---'s in the amino acid mutation column indicates the synonymous mutations.}
\item[a] \footnotesize{Genomic position as in \protect\cite{B.1.1.7} Table 1 and text above Table 1. Positions refer to SARS-CoV-2 sequence Wuhan-Hu-1 with the Genbank accession number ``MN908947.3".}
\item[b] \footnotesize{Frequencies of alleles have been computed
from the entire data set (reference) after 
multiple sequence alignment as described. Frequencies of
alleles at one locus have then been sorted as Allele 1 (major allele), Allele 2 (first minor allele), etc.}
\item[*]\footnotesize{The question mark ``?" indicates different nucleotides in the deletions.}
\item[c]\footnotesize{In time-sorted GISAID data
alleles at this locus have the opposite behavior than expected if the wild-type at this locus was C.
Using the same convention as the other loci 
we have take the mutation at this locus to be $T16176C$. 
The mutation at 23604 based on the whole prepared dataset also has a swapped mutation compared with that provided in the literature~\cite{B.1.1.7}.
}
\item[d] \footnotesize{
In time-sorted GISAID data the most
common allele at this locus is initially $C$ later overtaken by $G$.
However, the time course is very different from the rest of
the UK variant. Possibly this points 
to the use of another reference sequence
for this single mutation in gene $M$
in \protect\cite{B.1.1.7}.
Using the same convention as the other loci 
the mutation at this locus would be $C26801G$.
}
\item[e]\footnotesize{This locus is one of three annotated as $28280 GAT->CTA$ in \protect\cite{B.1.1.7} Table 1.}
\label{tab:B.1.1.7}
\end{tablenotes}
\end{threeparttable}
\end{table}

\subsection*{Definition of B.1.351 (``South Africa variant")} 
In this work we have used the definition of
SARS-CoV-2 Variant of Concern 202012/02
(B.1.351) as given in
``Technical briefing 6"~\cite{VoC-6}  
Table 4a.  This information with annotations is
given as Table~\ref{tab:B.1.351} below. 
The later Pangolin description of South Africa variant~\cite{PANGO_SA} is a subset of the above definition. The Defining Pangolin amino acid mutations are shown in the green cells of Table~\ref{tab:B.1.351}.

\begin{table}[!ht]
\centering
\caption{Defining mutations for variant B.1.351}
\begin{threeparttable}
\begin{tabular}{lllll}
gene sub-structure  & mutation locus \tnote{a} & amino acid mutation &  Allele 1 \tnote{b}  & Allele 2 \tnote{b}   \\

\hline
NSP2& C1059T & ORF 1ab: T265I     &  C & T\\
NSP3& G5230T & \cellcolor{green}ORF1ab: K1655N     &  G & T\quad\tnote{d}\\
NSP5& A10323G & ORF 1ab: K3353R     &  A & G \\
NSP6& 11288$\_$96 del & 3675-3677 del      &  -- & --\quad\tnote{e}\\
Spike& C21614T & S:L18F      &  C & T\quad\tnote{c} \\
Spike& A21801C & \cellcolor{green}S: D80A &  \cellcolor{lightgray}A &  \cellcolor{lightgray}N\quad\tnote{d} \\
Spike& A22206G & \cellcolor{green}S: D215G &   A & G\quad\tnote{d} \\
Spike& --  & 242-244del     &  -- & --\\
Spike& G22299T & R246I     &   \cellcolor{lightgray}G & \cellcolor{lightgray}N\quad\tnote{c} \\
Spike& G22813T & \cellcolor{green}S: K417N  &   G & T\quad\tnote{c,d}\\
Spike& G23012A & \cellcolor{green}SGF E484K &  G & A\quad\tnote{d} \\
Spike& A23063T & \cellcolor{green}S: N501Y     & A& T\quad\tnote{d,e} \\
Spike& C23664T & \cellcolor{green}S: A701V     &   C & T\quad\tnote{d}\\
ORF3a& G25563T & ORF3a: Q57H    &  G & T\\
ORF3a& C25904T & ORF3a: S171L    &  C & T\\
E& C26456T & \cellcolor{green}E: P71L     &  C & T\quad\tnote{d} \\
N& C28887T & \cellcolor{green}N: T205I     &  C & T\quad$^{\rm d}$\\
\hline
\end{tabular}
\begin{tablenotes}
\item[a] \footnotesize{Genomic position as in \protect\cite{VoC-6} Table 4a. Positions refer to SARS-CoV-2 sequence Wuhan-Hu-1 with the Genbank accession number ``MN908947.3".}
\item[b] \footnotesize{Frequencies of alleles have been computed
from the entire data set (reference) after 
multiple sequence alignment as described. Frequencies of
alleles at one locus have then been sorted as Allele 1 (major allele), Allele 2 (first minor allele), etc.}
\item[c]\footnotesize{ 
Annotated in \protect\cite{VoC-6} caption to Table 4a as acquisitions in subset of isolates within the lineage.}
\item[d]\footnotesize{
Annotated in \protect\cite{VoC-6} in Table 4b 
as ``PROBABLE"; at least 4 lineage defining non-synonymous changes called as alternate
base and all other positions either N or mixed base OR at least 5 of the 9
non-synonymous changes.}
\item[e]\footnotesize{
This mutation is also present in the UK variant,
compare Table~\protect\ref{tab:B.1.1.7}.}
\end{tablenotes}
\label{tab:B.1.351}
\end{threeparttable}
\end{table}

\subsection*{Definition of B.1.617.2 (``delta variant")} 
In this work we have used the definition of SARS-CoV-2 Variant of Concern B.1.617.2 as provided on the Nextstrain website \cite{Delta_variant}. The mutation information with annotations is given as Table~\ref{tab:B.1.617.2} below. The amino acid mutations with green background are presented in the Pangolin definition of delta variant~\cite{PANGO_delta}. They are all included in the third column of Table~\ref{tab:B.1.617.2}.
\begin{table}[!ht]
\centering
\caption{Defining mutations for Delta variant B.1.617.2}
\begin{threeparttable}
\begin{tabular}{lllll}
gene sub-structure  & mutation locus & amino acid mutation \tnote{a} &  Allele 1 \tnote{b}  & Allele 2 \tnote{b}  \\
\hline
ORF1b & C14408T   & ORF1b: P314L  &  \cellcolor{lightgray}T & \cellcolor{lightgray}C\quad\tnote{c}\\
ORF1b & C16466T  & ORF1b: P1000L  &  C & T\\
Spike & C21618G  & \cellcolor{green}S: T19R	   &  C & G \\
Spike & 22028-33 del & 156-157 del      &  -- & --\quad\tnote{d}\\
Spike & A22034G & S: R158G    &  A & \cellcolor{lightgray}N\quad\tnote{e} \\
Spike & T22917G  & \cellcolor{green}S:L452R   &  T & G\\
Spike & C22995A & \cellcolor{green}S: T478K    &  C & A \\
Spike & A23403G & D614G    &  \cellcolor{lightgray}G & \cellcolor{lightgray}A\quad\tnote{c} \\
Spike & C23604G & \cellcolor{green}S: P681R    &  \cellcolor{lightgray}A & \cellcolor{lightgray}C\quad\tnote{f}\\
Spike & G24410A & \cellcolor{green}S: D950N    &  G & A \\
ORF3a & C25469T & \cellcolor{green}ORF3a: S26L     &  C & T \\
M     & T26767C & \cellcolor{green}M: I82T     &  T & C \\
ORF7a & T27638C & \cellcolor{green}ORF7a: V82A     &  T & C \\
ORF7a & C27752T & \cellcolor{green}ORF7a: T120I    &  C & T \\
N     & A28461G & \cellcolor{green}N: D63G     &  A & G\\
N     & G28881T & \cellcolor{green}N: R203M    &  \cellcolor{lightgray}A & \cellcolor{lightgray}G\quad\tnote{f}\\

N     & G29402T & \cellcolor{green}N: D377Y    &  G & T \\
\hline
\end{tabular}
\begin{tablenotes}
\item[a] \footnotesize{Amino acid mutations provided on the website~\cite{Delta_variant} from corresponding protein. The Mutation loci in the second column refer to SARS-CoV-2 sequence Wuhan-Hu-1 with the Genbank accession number ``MN908947.3". The green ones are listed in the Pangolin description.}
\item[b] \footnotesize{Frequencies of alleles have been computed
from the entire data set. Frequencies of
alleles at one locus have then been sorted as Allele 1 (major allele), Allele 2 (first minor allele), etc.}
\item[c]\footnotesize{The wild type and the second most common alleles in the definition of B.1.617.2 have been swapped.}
\item[d]\footnotesize{These deletions are still there.}
\item[e]\footnotesize{
This mutation based on the whole dataset became deletion now.}
\item[f]\footnotesize{
The wild type in these mutations based on the whole dataset has absent in the definition (second column). The wild type in the definition  became the second most common allele based on the dataset.}
\end{tablenotes}
\end{threeparttable}
\label{tab:B.1.617.2}
\end{table}

\section{Results}\label{sec:results}
The GISAID repository holds a large collection of whole-genome sequences~\cite{GISAID}. We have downloaded high quality sequences up to the end of July, 2021.
All collected genomes are aligned in an MSA table. The frequency of a given allele at a given locus is then how many times that allele is found in the corresponding column in the table divided by the number of rows of the table. In this way we determine the frequency of a mutation at one locus as a function of time. Similarly, we determine the joint frequency of a set of mutations as the number of times all these mutations are found in the same row in the table, divided by the number of rows in the table.  

\subsection{Temporal behavior of B.1.1.7 variant}

\begin{figure}[!ht]
\centering
\includegraphics[width=0.80\textwidth]{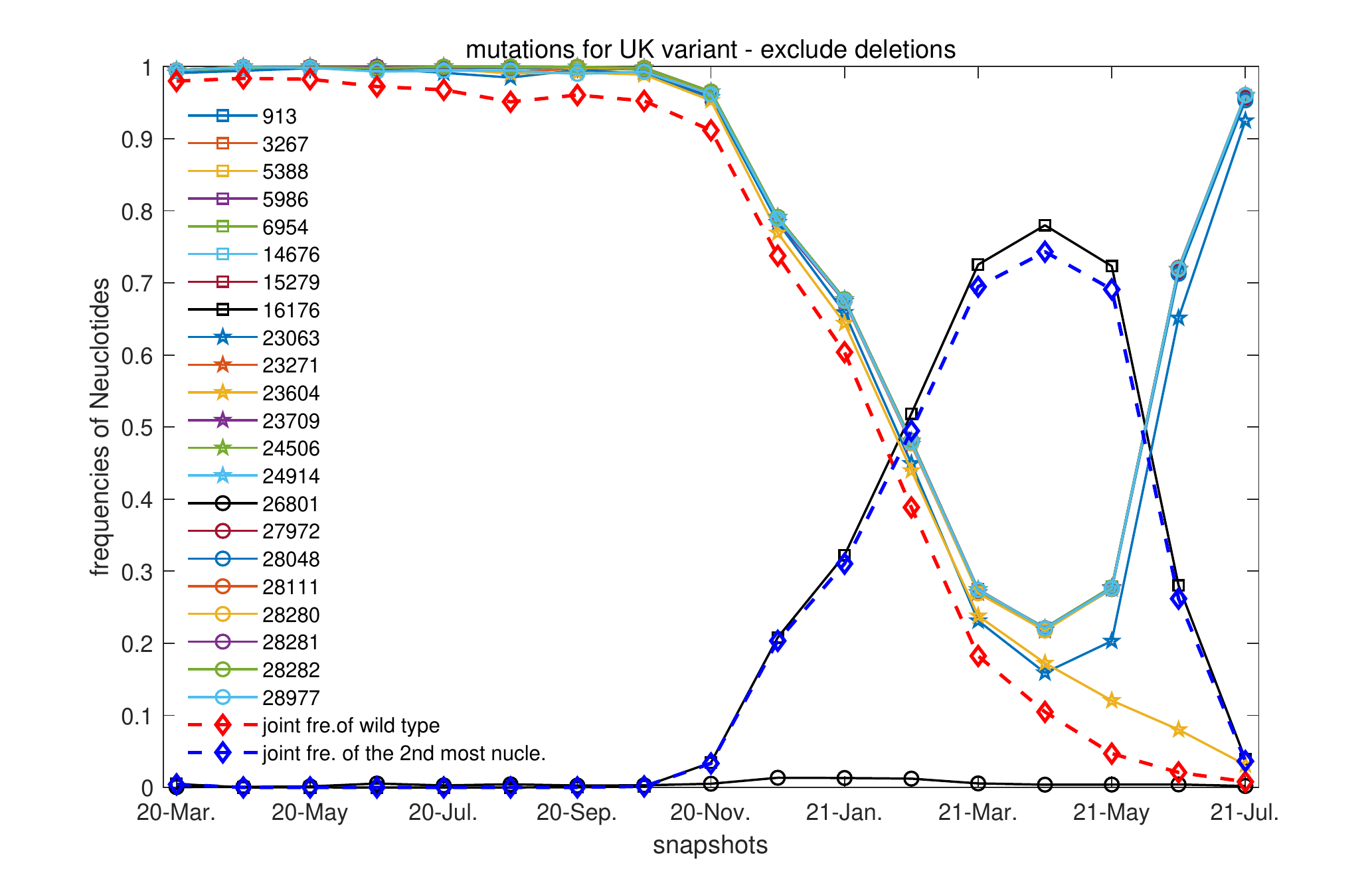}
\caption{Individual frequency of major allele (solid marked-lines) and the joint frequency (dashed diamond-lines) at the defining loci for B.1.1.7 over time as determined from GISAID. The red dashed line is for the joint frequency of the wild type while the blue one for the second most common allele of the B.1.1.7 variant. 21 out of 23 mutations listed for the UK variant report have similar temporal patterns, except the 26801 locus in M gene and the mutation of $C16176T$ in NSP12. Two outliers are not included in the joint frequency analysis.}
\label{fig:frequeces_of_loci_monthly_UK}
\end{figure}

The first report from Public Health England defining B.1.1.7 as a Variant of Concern lists 17 non-synonymous mutations (including deletions) and six synonymous mutations ~\cite{B.1.1.7}, see  Table~\ref{tab:B.1.1.7}.
The time series of the frequencies of these mutations is shown in  Fig.~\ref{fig:frequeces_of_loci_monthly_UK}. Of the 23 mutations, 21 have a similar time course, C16176T has the precise opposite time course, and T26801C an unrelated time course. In the following we have assumed that C16176T is a mis-labelling, and that this mutation in fact is T16176C. We have further assumed that T26801C, a synonymous mutation in the M gene, pertains mostly to another clone or to another reference sequence. In the following we have not retained data from this locus. In Fig.~\ref{fig:frequeces_of_loci_monthly_UK} we also show the time course of the joint frequency of the 22 retained mutations for this variant.

\begin{figure}[!ht]
\centering
\includegraphics[width=0.8\textwidth]{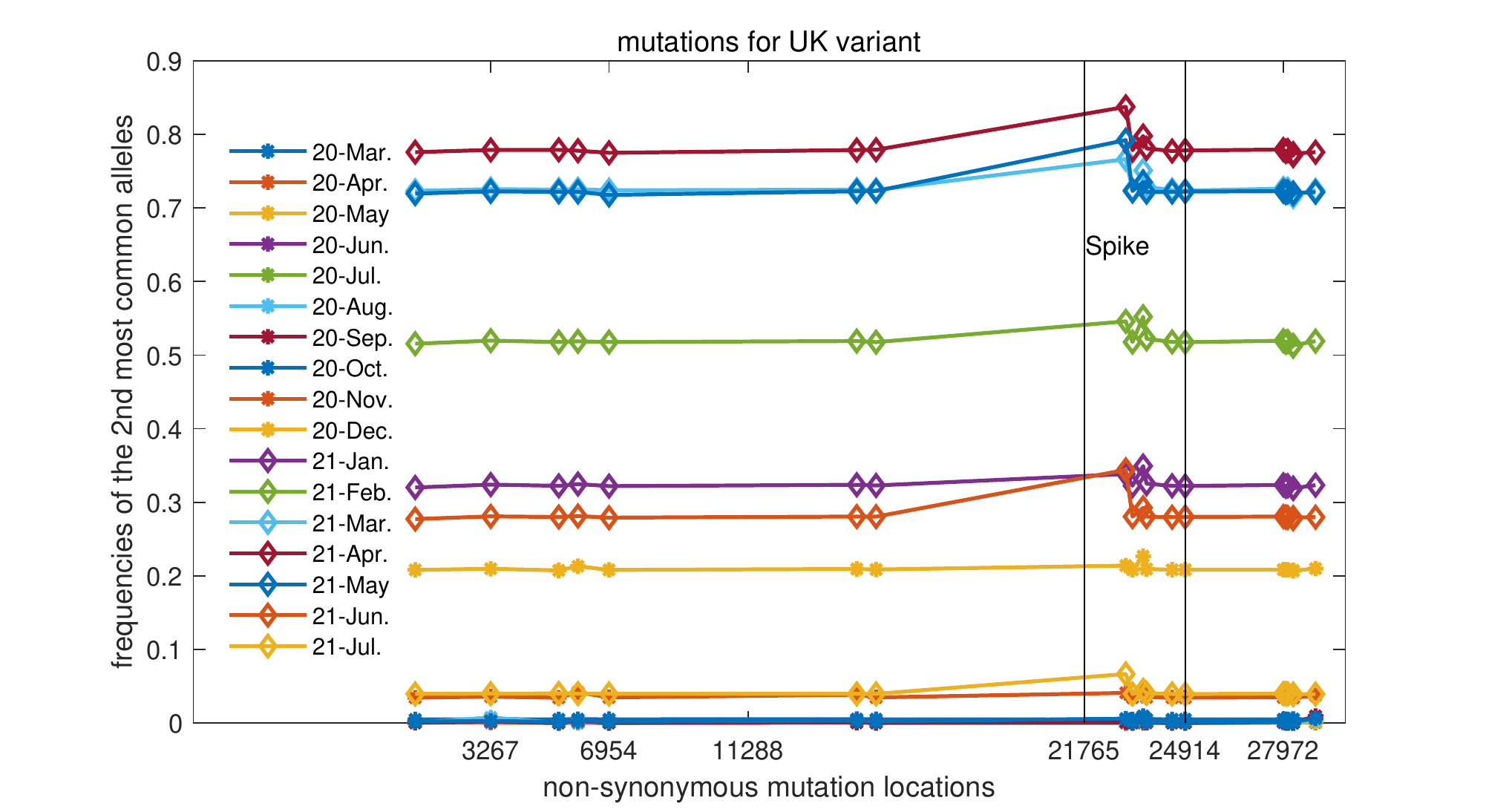}
\caption{Frequency of second most common allele at the defining loci for B.1.1.7 over time as determined from GISAID. X-axis gives genomic positions. The 26801 locus in M gene is not included compared with Fig.1. The average distance between the 22 retained mutations is about $1,500$ bp, but some such as $C23604A$ and $C23709T$ ($P681H$ and $T716I$ in Spike) lie closer. The  mutations happened right to Spike gene are close to each other also.}
\label{fig:frequeces_of_loci_2nd_UK}
\end{figure}

The frequencies of the 22 retained mutations for the UK-variant increase after late summer / early autumn 2020, see Fig. \ref{fig:frequeces_of_loci_2nd_UK}. The lines in this figure connect frequencies of the second most common allele (first minor allele) within the same month of sampling time in the GISAID data. With one exception (16176, discussed above) this second most common allele agrees with the mutation at this locus as given in~\cite{B.1.1.7}.

The growth of the first minor allele of the UK-variant is uneven across the SARS-CoV-2 genome. In a first phase (early 2020-November 2020), the frequency of the HV 69-70 deletion in Spike (21765-21770) is noticeably higher than the other mutations defining B.1.1.7. This is consistent with this mutation initially being present also in clones unrelated to B.1.1.7. As time progresses, the relative difference between the frequency at this locus and the frequencies at the other loci decreases. Starting in December 2020 for C23604A (P681H in Spike), January 2021 for A23063T (N501Y in Spike) and February 2021 for the deletion 11288-11296 in NSP6 the frequencies of these three mutations begin to differ in frequency, and these differences increase up to May 2021. The joint frequency of the mutations hence progressively deviates from that of the single mutations in the spring of 2021.

\subsection{Temporal behavior of B.1.351 variant}
\begin{figure}[!ht]
\centering
\includegraphics[width=0.80\textwidth]{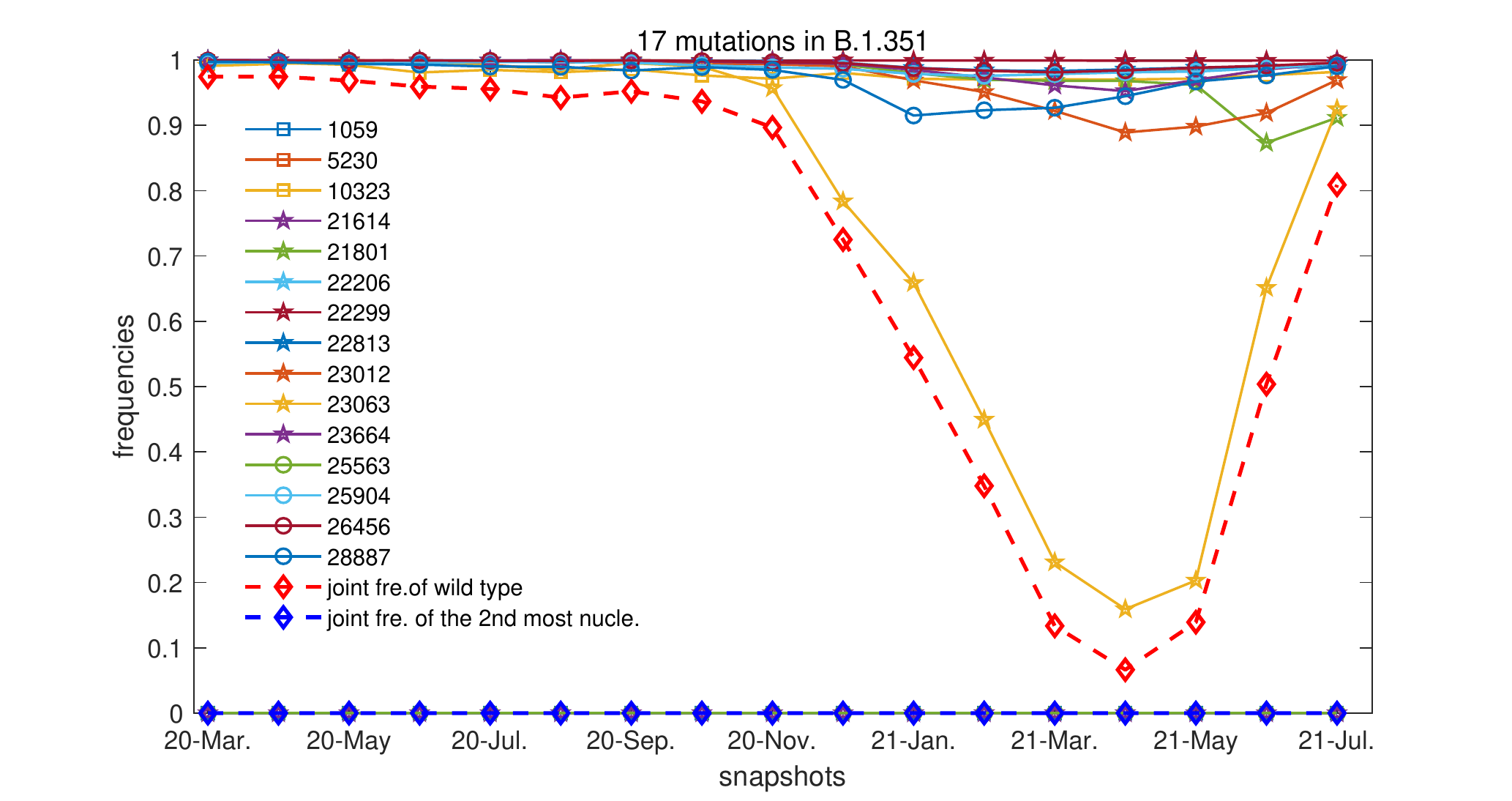}
\caption{Individual and joint frequency of major allele at the defining loci for B.1.351 over time as determined from GISAID. They are plotted in solid marked-lines and dashed diamond-lines respectively. Three out of $17$ mutations listed for the South Africa variant (one is 1059, the other two 21614, 25563 are covered by the diamond lines) display different dynamics and have been excluded from the following analysis. Of the others, two mutations shared with B.1.1.7 increase to large frequencies: the $3675-3677$ deletion ($11288-11296$) in NSP6 and the $N501Y$ mutation ($A23063T$) in Spike (shown in pentagonal lines). The remaining mutations reach about the 2\% level and are discussed in text.}
\label{fig:frequeces_of_loci_monthly_SA}
\end{figure}

The definition of B.1.351 given by Public Health England in February 2021, lists 17 non-synonymous mutations (including deletions) out of which 9 in Spike~\cite{VoC-6},
see Table~\ref{tab:B.1.351}.
In later descriptions, such as in Pangolin, only a fraction of these mutations has been retained, 
see the cells in Tab.~\ref{tab:B.1.351} with green background.
Of the 17 mutations listed in the first PHE definition, two are also listed in B.1.1.7. As this variant reached higher prevalence world-wide, frequencies of mutations at those loci followed that course and we have not retained them for B.1.351. Three other mutations listed in the first PHE definition appeared much before B.1.351 was defined and have an unrelated time course, see 
three appeared much before this variant was
defined and have an unrelated time course,
see Fig.~\ref{fig:frequeces_of_loci_monthly_SA}.
In the following we have assumed 
that these three
mutations, $C1059T$ ($T265I$ in $NSP2$),
$C21614T$ ($L18F$ in Spike)
and $G25563T$ ($Q57H$
in $ORF3a$) also mostly pertain to other 
clones and/or to another reference sequence.
We have not retained data 
from these loci.

\begin{figure}[!ht]
\centering
\includegraphics[width=0.70\textwidth]{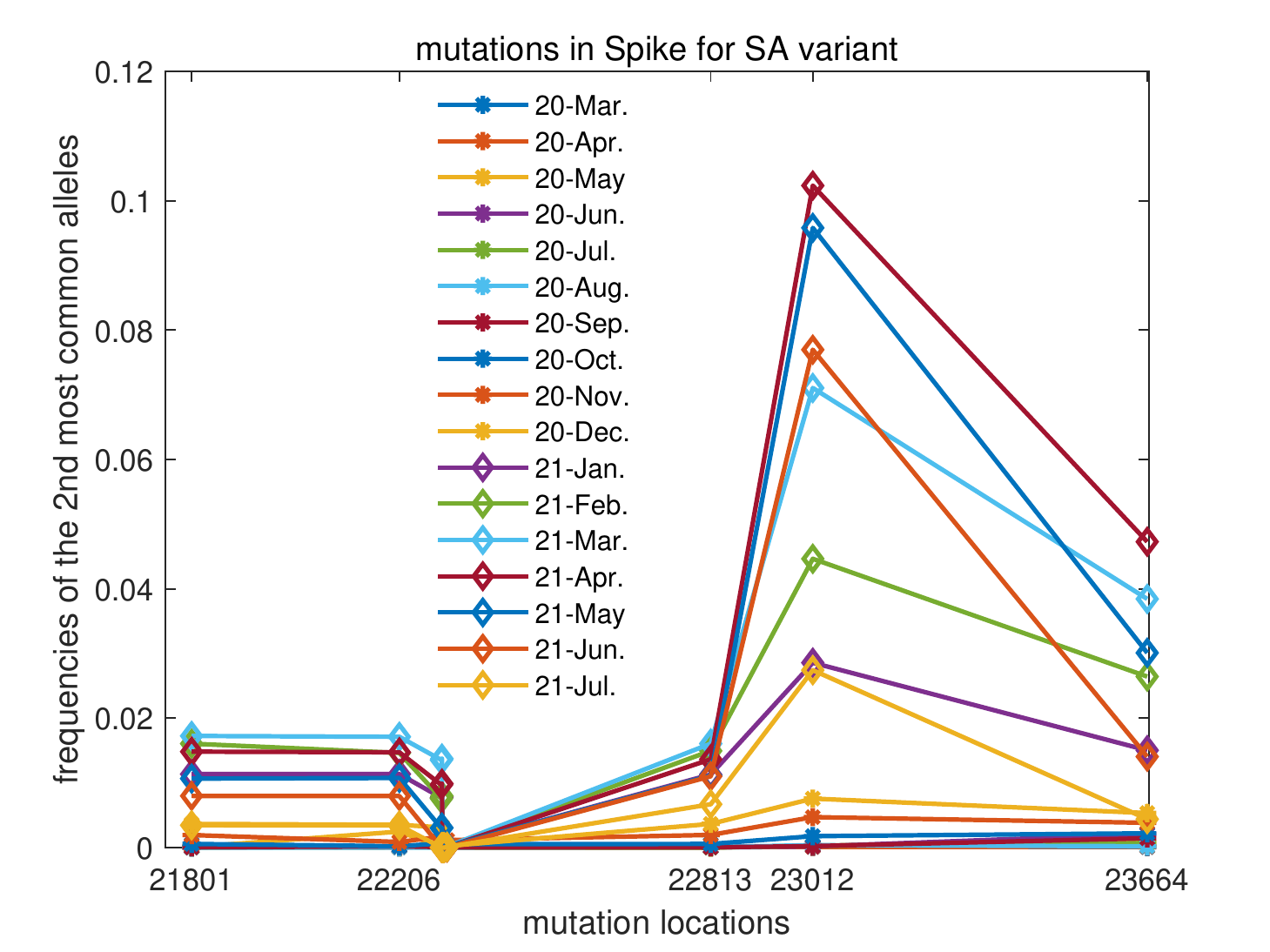}
\caption{ Frequency of second most common allele at the defining loci in Spike for B.1.351 over time as determined from GISAID. X-axis gives genomic positions. 
The mutation at $23012$ is SGF $E484K$ in Spike $S2$. The mutation $N501Y$ ($A23063T$) is also defined for B.1.1.7 and is not shown, see instead Fig.~\protect\ref{fig:frequeces_of_loci_2nd_UK}.
}
\label{fig:frequeces_of_loci_monthly_2nd_SA}
\end{figure}

To display the frequencies of mutations in B.1.351, we have chosen to focus on mutations in Spike, except N501Y in Spike ($A23063T$) which is shared with the UK variant, see Fig. ~\ref{fig:frequeces_of_loci_monthly_2nd_SA}.
The growth of these mutations in B.1.351 in Spike are as follows. From the beginning of Spike up and including the 242-244 deletion (three loci) there is a roughly even growth up to approximately the 2\% level up to March 2021. April 2021 is still even while May 2021 is more uneven with A21801C up and $A22206G$ down. Immediately to the right of this set there is a sharp drop in frequency so that very few sequences carry the $R246I$ mutation. This mutation has now been excluded from the defining mutations of B.1.351, together with several other mutations originally listed~\cite{PANGO_lineages}. 
Further to the right, $K417N$ ($G22813T$) follows approximately the pattern of the 242-244 deletion while $A701V$ ($C23664T$) appears to grow about twice as fast and reaches approximately the 5\% level in May 2021. $E484K$ ($G23012A$) on the other hand follows an erratic trajectory peaking at above 1\% in August 2020, falling back to below 1\% in November 2020, and then increasing again up to 9\% in April 2021 and then falling back to 6\% again in May 2021. This reflects the fact that this mutation is shared with the $P1$ and other variants~\cite{PANGO_lineages}.

\subsection{Temporal behavior of B.1.617.2 variant}

\begin{figure}[!ht]
\centering
\includegraphics[width=0.80\textwidth]{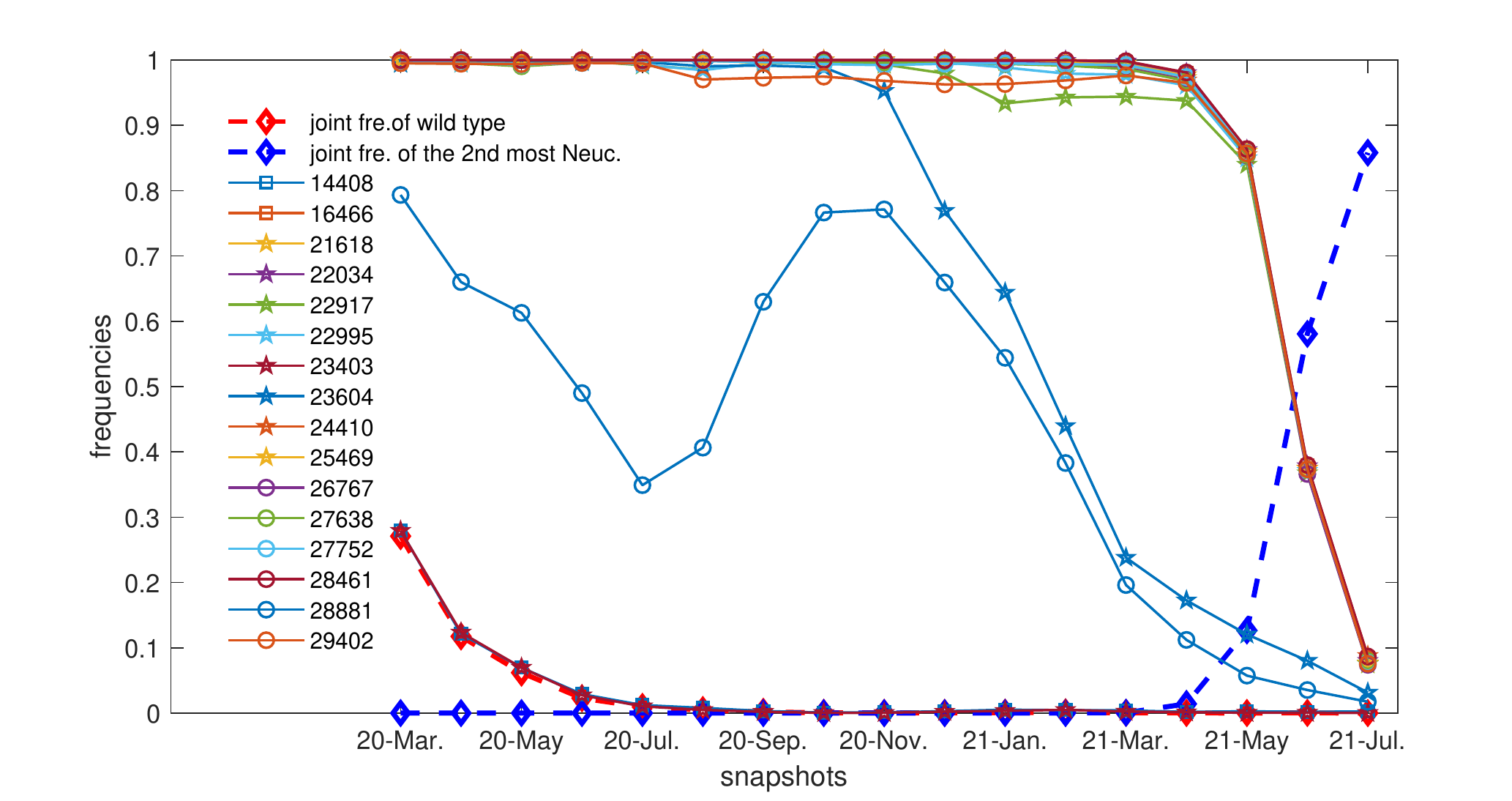}
\caption{Individual and joint frequency of major allele at the defining loci for B.1.617.2 over time as determined from GISAID. They are plotted in solid marked-lines and dashed diamond-lines respectively. The deletions are not shown here. The allele frequency of the Wild type at most mutated locus went down around April while 23604 around the end of 2020. 28881 is a well-known locus that has heavy variations. 14408 and 23403 have a similar behavior with that of the joint frequency for the wild type.}
\label{fig:frequeces_of_loci_monthly_Delta}
\end{figure}

The B.1.617.2 variant also known as ``delta variant", was first identified in India in December 2020. It has contributed to the widespread cases of COVID-19 all over the world today. The definition of this variant we used here is listed on the Nextstrain website, including 17 non-synonymous (including deletions) out of which 8 in Spike~\cite{Delta_variant}. In the later Pangolin version of delta variant~\cite{PANGO_delta}, two mutations from ORF1b and 3 in Spike are not included, as shown in cells of Table~\ref{tab:B.1.617.2} with green background.
There are mainly three kinds of temporal behavior of mutations for this variant at the early stage of evolution, see Fig.\ref{fig:frequeces_of_loci_monthly_Delta}. The allele frequencies of most wild types start decreasing around April 2021 while that of 23604 and 28881 went down around the end of year 2020. The temporal allele frequencies of 14408 and 23403 have a similar behavior with that the joint frequency of the wild type. The frequencies of all listed alleles at these four locations are computed from the whole dataset we used in this work. They are sorted and the first and second prevalent alleles are labelled as Allele 1 and Allele 2 respectively in Table~\ref{tab:B.1.617.2}. The above 4 mutations are marked with lightgray, which indicates the mutations calculated from dataset at these loci are different from the definitions. For instance, $C14408T$ became $T14408C$ and $A23403G$ became $G23403A$. For the $C23604G$ mutation, the original wild type became the second most common allele while the domain one came from the minor alleles which is not included in the definition. The same situation also holds for the mutation of $G28881T$.
The joint frequency of the second most common alleles as shown in the diamond lines starts going up around April 2021 reached at around 90\%. This indicates why the delta variant has been detected everywhere nowadays.

\begin{figure}[!ht]
\centering
\includegraphics[width=0.80\textwidth]{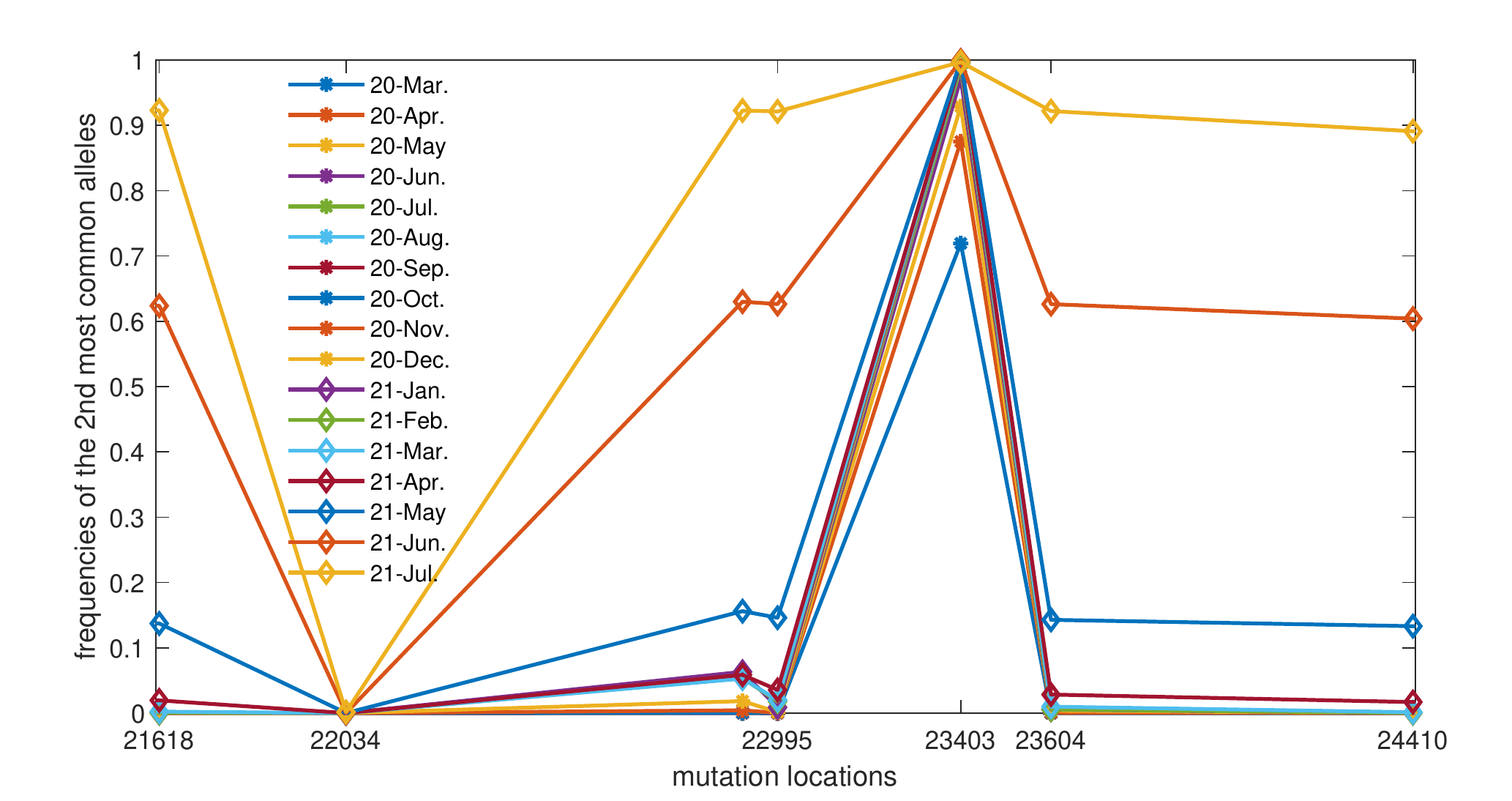}
\caption{Individual frequency of second most common allele at the defining loci in Spike for B.1.617.2 over time as determined from GISAID. X-axis gives genomic positions. Most of the mutations grow up with time goes on while this never happen at 22304. This is consistent with the annotated mutations in Table~\ref{tab:B.1.617.2} for this locus, the ``Allele 2" became ``N" till now instead of ``G" shown in the definition.
}
\label{fig:frequeces_of_loci_monthly_2nd_Delta}
\end{figure}

Similarly to the analysis of the previous two variants, the frequencies of the second most common alleles for delta variant are also computed, as shown in Fig. \ref{fig:frequeces_of_loci_monthly_2nd_Delta}. Here only mutations in Spike are shown. Obviously, most of the allele frequencies growing up with time goes on. However, there is one outlier $22034$ whose frequency never goes up and around zero. This is consistent with the corresponding mutation in Table \ref{tab:B.1.617.2}. Based on the computation on the dataset, the second most common allele at this location is `N', which means the mutation at this locus may become ``deletion" instead of the $A22034G$ in the definition.

\section{Discussion}\label{sec:discussion}
One conclusion of this work is that it cannot be the case that the 
three variants, as originally defined, have grown as large clones. 
No sophisticated statistical analysis is required to reach this conclusion. 
It is enough to plot the frequencies of single locus mutations and eyeball the time series curves.
In this work we have used almost two million whole-genome SARS-CoV-2 sequences from GISAID. 
The method of analyzing allele frequencies and plotting the data stratified by time is a simple and efficient way of visualizing the dynamics of variants when there is such abundant data available on the genomic level.
With the growing availability of genomic sequencing and the possibility of sequencing samples in disease outbreaks one can expect that this will soon become standard. Not only for serious
pandemics like COVID19, but also for milder infectious diseases such as seasonal influenza. 

More in detail, SARS-CoV-2 variant alpha originally grew as one clone -- with the exception of 
two likely mis-labelled mutations and the $HV 69-70$ deletion, which diverged into separate clones as the variant accumulated novel mutations. Indeed, over time the list of defining mutations for B.1.1.7 was adjusted and the synonymous mutations removed.  It is evident that SARS-CoV-2 variant beta originally contained several mis-classified mutations. Most notably 21614 in the Spike protein, $1059$ in NSP2 and $25563$ in ORF3a, all of which subsequently were removed from the list of defining mutations~\cite{B.1.1.7, VoC-6}. Regardless of these mutations it is clear that this variant does not grow as one large clone, but rather consists of multiple variants. 

For delta variant, except the well known  mutation located at 23604 and 28881 which has already been detected by the pairwise analysis based on the dataset till the beginning of August 2020~\cite{Zeng2020}, the temporal frequencies in Fig.~\ref{fig:frequeces_of_loci_monthly_Delta} clearly show two different patterns which may indicates the non-unique clone in the evolution. Meanwhile, the 
silent mutation at $22034$ locus in Fig.~\ref{fig:frequeces_of_loci_monthly_2nd_Delta} probably shows a problematic definition of mutation $A22034G$ at this position. 
This may be why this mutation is not included in the Pangolin version of delta variant definition.

This instability of clones is supported by recent observations pointing towards the emergence of multiple lineages of SARS-CoV-2 within the same individual
\cite{Avanzato2020,Baang2021,Choi2020,Hensley2021,Kemp2021}. 
In all cases the patients had prolonged viremia and received convalescent plasma treatment and/or monoclonal antibody therapy. Treatment with convalescent plasma or monoclonal antibodies applies selection pressure on a viral population within the host that may drive the emergence of antibody resistant clones. Also, the large number of viral genomes present simultaneously in a single patient enable opportunities for within host recombination. The phenotypic effects of all described mutations in the spike protein of SARS-CoV-2 are just beginning to be unraveled. For example, the N501Y substitution increases the affinity for ACE2 binding~\cite{Gu2020}. Also, compensatory mutations have been described as in the case for the E484K substitution in combination with del69-70, where a reduction in antibody sensitivity is compensated with increased infectivity. Specific mutations that confer advantages, in the form of increased infectivity or antibody escape, will increase in frequency as they are shared with novel variants that either emerge from the original or emerge in unrelated variants. This is illustrated for E484K of B.1.351 which increases more rapidly than the other defining mutations for this variant. The E484K mutation is also present in for example the P1 variant and a sub-variant of the B.1.1.7~\cite{VoC-6}.

On a more general and speculative note, Coronaviruses, the larger family to which SARS-CoV-2 belongs,
in general exhibit a large amount of recombination~\cite{LaiCavanagh1997,Graham2010}. 
Large-scale recombination would be important in the 
COVID19 pandemic for several reasons.
First it increases the resiliance of the viral population 
against hostile agents. Beneficial (to the virus) changes can spread faster
and more reliably throughout the population.
Second it leads to form of evolution optimizing fitness and less impacted by traits inherited by chance: while a clone replicating asexually 
will likely have points of weakness,
in a recombining population such errors are shared around and eliminated.
Third, substantial amount of recombination 
is a confounder for phylogentic reconstruction.
Crudely put, phylogenetic trees
reconstructed from population-wide sequence data
may not reflect the actual evolution in such
populations, an issue which has been discussed in
bacterial phylogenetics since some time
\cite{Falush2003,Dixit2017,Sakoparnig2021}.

Overall, our discussion underlines the usefulness of a time series analysis when handling data on systems as complex as an evolving genome may be. This is almost a truism from the point of view of complexity scientist: in the context of network science, for instance, a whole new branch of the field has developed with a focus on temporal dimension \cite{Holme2012,Holme-2019}. Indeed, projecting out the temporal dimension in a single static "snapshot" of the systems on one hand allows rapid analysis and computations on the data but on the other hand may be harmful when the dynamical time scales
at stake are not separable or unknown, which is almost always the case for evolutionary and epidemiological models (mutation rate of the genome, recombination rate, rate of diffusion and so on). We thus suggest that the use of time series analysis can be a useful complement to more automated and sophisticated forms of data analysis, whose assumptions may hide important and perhaps dominant phenomena.


\bibliographystyle{iopart-num}
\bibliography{Complexity}

\section*{Acknowledgments}
We thank Richard Neher for comments on 
a first version of the MS, and for pointing
out that in the annotation used by
nextstrain, $C26801G$ is counted in clade
20E (EU1).
The work of HLZ was sponsored by National Natural Science Foundation of China (11705097).
The work of EA
was supported by 
the Swedish Research Council grant 2020-04980.

\end{document}